\documentclass[%
reprint,
superscriptaddress,
 amsmath,amssymb, aps,
floatfix,
]{revtex4-1}
\usepackage{amsmath}
\usepackage{marvosym}
\usepackage{graphicx,subfigure}         
\usepackage{amsmath} 
\usepackage{amssymb,stmaryrd}   
\usepackage{color}
\usepackage{wrapfig}
\usepackage{mathtools}
\usepackage{tikz}  
\usetikzlibrary{arrows,shapes,trees,positioning}  
\usepackage{comment} 
\usepackage{units}
\usepackage{siunitx}

\usepackage{appendix}
\usepackage{hyperref}
\usepackage{textcomp}
\hypersetup{
    colorlinks=true,
    linkcolor=blue,
    citecolor=blue,
    urlcolor=blue,
}
\usepackage{ragged2e}
\usepackage{ulem}

\begin{document}

\preprint{APS/123-QED}

\title{{Element resolved evidence of superdiffusive spin current arising from ultrafast demagnetization process\\
}}

\author{R. Gupta}
\email{rahul.gupta@angstrom.uu.se}
\affiliation{Department of Materials Science and Engineering, Uppsala University, Box 35, SE-751 03, Uppsala, Sweden}
\affiliation{Department of Physics and Astronomy, Uppsala University, Box 516, SE-751 20, Uppsala, Sweden}

\author{F. Cosco}
\affiliation{Department of Physics and Astronomy, Uppsala University, Box 516, SE-751 20, Uppsala, Sweden}

\author{R. S. Malik}
\affiliation{Department of Physics and Astronomy, Uppsala University, Box 516, SE-751 20, Uppsala, Sweden}

\author{X. Chen}
\affiliation{Department of Physics and Astronomy, Uppsala University, Box 516, SE-751 20, Uppsala, Sweden}
\affiliation{Department of Physics and Shenzhen Institute for Quantum Science \& Engineering, Southern University of Science and Technology, Shenzhen, Guangdong 518055, China}

\author{S. Saha}
\affiliation{Department of Physics and Astronomy, Uppsala University, Box 516, SE-751 20, Uppsala, Sweden}
\affiliation{Department of Physics, Ashoka University, 131029 Hariyana, India}

\author{A. Ghosh}
\affiliation{Department of Physics and Astronomy, Uppsala University, Box 516, SE-751 20, Uppsala, Sweden}
\affiliation{MAX IV Laboratory, Lund University, SE-22 100, Lund, Sweden}

\author{T. Pohlmann}
\affiliation{Deutsches Elektronen-Synchrotron (DESY), Notkestrasse 85, D-22607 Hamburg, Germany}

\author{J. R. L. Mardegan}
\affiliation{Deutsches Elektronen-Synchrotron (DESY), Notkestrasse 85, D-22607 Hamburg, Germany}

\author{S. Francoual}
\affiliation{Deutsches Elektronen-Synchrotron (DESY), Notkestrasse 85, D-22607 Hamburg, Germany}

\author{R. Stefanuik}
\affiliation{Department of Physics and Astronomy, Uppsala University, Box 516, SE-751 20, Uppsala, Sweden}

\author{J. S{\"o}derstr{\"o}m}
\affiliation{Department of Physics and Astronomy, Uppsala University, Box 516, SE-751 20, Uppsala, Sweden}

\author{B. Sanyal}
\affiliation{Department of Physics and Astronomy, Uppsala University, Box 516, SE-751 20, Uppsala, Sweden}

\author{O. Karis}
\affiliation{Department of Physics and Astronomy, Uppsala University, Box 516, SE-751 20, Uppsala, Sweden}

\author{P. Svedlindh}
\affiliation{Department of Materials Science and Engineering, Uppsala University, Box 35, SE-751 03, Uppsala, Sweden}

\author{P. M. Oppeneer}
\affiliation{Department of Physics and Astronomy, Uppsala University, Box 516, SE-751 20, Uppsala, Sweden}

\author{R. Knut}
\email{ronny.knut@physics.uu.se}
\affiliation{Department of Physics and Astronomy, Uppsala University, Box 516, SE-751 20, Uppsala, Sweden}

\date{\today}

\begin{abstract}
Using element-specific measurements of the ultrafast demagnetization of Ru/Fe$_{65}$Co$_{35}$ hetero-structures, we show that Ru can exhibit a significant magnetic contrast (3\% asymmetry) resulting from ultrafast spin currents emanating from the demagnetization process of the FeCo layer. We use this magnetic contrast to investigate how superdiffusive spin currents are affected by the doping of heavy elements in the FeCo layer. We find that the spin currents are strongly suppressed,  and that the recovery process in Ru slows down, by Re doping. This is in accordance with a change in interface reflectivity of spin currents as found by the superdiffusive spin transport model. 
\end{abstract}

\maketitle

 In 1996 it was found by Beaurepaire \textit{et al.}\ \cite{Beaurepaire_1996}  that a ferromagnetic Ni film can be demagnetized on a subpicosecond time scale by femtosecond (fs) laser pulse excitations. The possibility to optically manipulate the magnetization on such fast time scales paves the way for terahertz (THz) operation of spintronic devices \cite{zhang2020ultrafast}. Moreover, ultrafast demagnetization of Ni results in THz emission being proportional to the second time-derivative of magnetization \cite{beaurepaire2004coherent}. Besides technological relevance, research in this area has led to discoveries of novel phenomena like all-optical magnetization switching  \cite{Stanciu_2007,Gorchon_2017}, transfer of THz spin currents~\cite{Rudolf_2012,kampfrath2013terahertz,Schellekens_2014} and the optical inter-site spin transfer effect  \cite{dewhurst2018laser,Willems_2020,Tengdin_2020}. 
For the last two decades, ultrafast magnetization dynamics of magnetic materials has been probed with fs laser pulses using time-resolved techniques such as high harmonic generation (HHG)~\cite{mathias2012probing,Turgut_2013}, magnetic circular dichroism~\cite{Stamn_2007}, and spin-polarized two-photon photoemission~\cite{Schmidt_2010}. Several microscopic mechanisms have been  proposed to explain the ultrafast dynamics including  electron-photon mediated spin-flip scattering~\cite{Koopmans_2005,Koopmans_2010,Steiauf_2009}, electron-electron scattering~\cite{Krauss_2009}, electron-magnon scattering~\cite{Carpene_2008,PhysRevResearch.2.013180}, and direct transfer of angular momentum from photon to electron mediated by spin-orbit coupling~\cite{DallaLonga_2007,Zhang_2000}. Malinowski \textit{et al.}~\cite{Malinowski_2008} first showed that laser-induced spin transport can enhance the demagnetization in metallic heterostructures. Superdiffusive spin transport \cite{Battiato_2010} was proposed to play an important role in the laser-induced demagnetization process of metallic magnetic films~\cite{Rudolf_2012,Eschenlohr_2013,Bergeard2016,Hofherr2017}. However, other mechanisms to generate fast spin currents such as spin pumping were also proposed  \cite{Choi2015,Remy2020,Beens2020} and the origin  of optically induced spin currents is still being debated \cite{Lichtenberg2022}. Spin-current injection through spin pumping \cite{Choi2015,Beens2020} should be favored by a stronger damping, but, although the latter can be engineered e.g.\ by doping of the ferromagnetic layer, such investigations are scarce.
 
Kampfrath \textit{et al.}\ investigated the THz emission caused by the ultrafast laser-induced demagnetization of an Fe layer adjacent to nonmagnetic Ru and Au capping layers \cite{kampfrath2013terahertz}. The measured temporal behavior of the THz spin currents was very different between the Ru and Au layers, which was explained by the different absorption of superdiffusive spin currents in the Ru and Au layers. In the case of Au, there are no empty $d$-states and the hot electrons injected into the $sp$ states with high electron-band velocity will therefore be quickly reflected back into the magnetic layer. In the case of Ru, the $d$-band states with lower band velocities will  transiently trap the injected spins and thus delay the spin current. However, as only the THz emission was measured, there was no \textit{de facto} evidence of a superdiffusive spin current in the nonmagnetic Ru layer.

In this work, we investigate the relation between ultrafast demagnetization and spin current generation as well as
effects of doping and choice of capping layer
by using the element-specific transverse magneto-optic Kerr effect (T-MOKE) and superdiffusive spin current simulations. Ru/Fe$_{65}$Co$_{35}$/Ru and Cu/Fe$_{65}$Co$_{35}$/Cu heterostructures are utilized as model systems, where rhenium (Re) is used as dopant in Fe$_{65}$Co$_{35}$ to modify the spin dynamics (cf.\ Fig.\ S1 in Supplemental Materials (SM) \cite{SM}).
The ultrafast spin dynamics is shown to be strongly dependent on the Re doping concentration. A transient spin accumulation, that is significantly reduced with increasing Re doping, is evident in the Ru layer. This decreased spin injection with Re doping can partly be assigned to a decreasing sample magnetization, and potentially a decreased spin transport due to scattering at Re sites. However, our simulations show that an increased electron reflectivity at the Ru interface, due to Re doping, results in a good qualitative correspondence to the  temporal behaviour of the Ru spin accumulation found experimentally.

The Re doped (0, 3, 6, and 12.6 at.\%) Fe$_{65}$Co$_{35}$ thin films were grown on SiO$_2$/Si substrate at room temperature using DC sputtering. Layers of Ru(3 nm) and Cu(5 nm) were used as seed and capping layers. The nominal thickness of Fe$_{65}$Co$_{35}$ was 20 nm, which was verified by x-ray reflectometry. All characterization details can be found in section I of the SM \cite{SM} and in Refs.\ \cite{akansel2018effect,rahulenginner,serkanenhaced}. 

The element-specific ultrafast magnetization dynamics of Fe$_{65}$Co$_{35}$ interfaced with Ru and Cu films as function of Re doping was studied using a table-top HHG setup \cite{mathias2012probing,helios,somnath_01,Robert2018}. The table-top HHG setup uses the T-MOKE to measure the magnetic asymmetry of the signal (cf. section I, SM \cite{SM}). In the T-MOKE spectrometer, the laser is a chirped pulse amplification system running at a central wavelength of 800 nm with a maximum energy of 2 mJ per pulse and a pulse length of 35 fs. The sample is magnetized, using a magnetic field strength of $\pm$ 80 mT, perpendicularly to the plane of incidence of the incoming $p$-polarized light \cite{somnath_01}. The reflected light from the sample is recorded with a CCD camera. In order to obtain a magnetic contrast, the change in reflected intensity at the absorption edges of the different elements in the heterostructure was  recorded for opposite directions of the magnetization. The resulting photon-energy dependent magnetic asymmetry $A(E)$ is the normalized difference of the reflected intensities for the two magnetization directions \cite{mathias2012probing,Jana2020} and is defined as follows,
\begin{equation}
A (E) =\frac{I_\mathrm{+}(E)- I_\mathrm{-}(E)}{I_\mathrm{+}(E)+I_\mathrm{-}(E)},
\end{equation}
where $I_{+}(E)$ and $I_{-}(E)$ are the reflected intensities measured for the two different magnetization directions. This asymmetry is proportional to the magnetization of the sample as discussed in previous studies \cite{mathias2012probing,Mathias2013164,chan2012ultrafast,Jana2020}.

\begin{figure}
    \centering
    \includegraphics[width = 9cm]{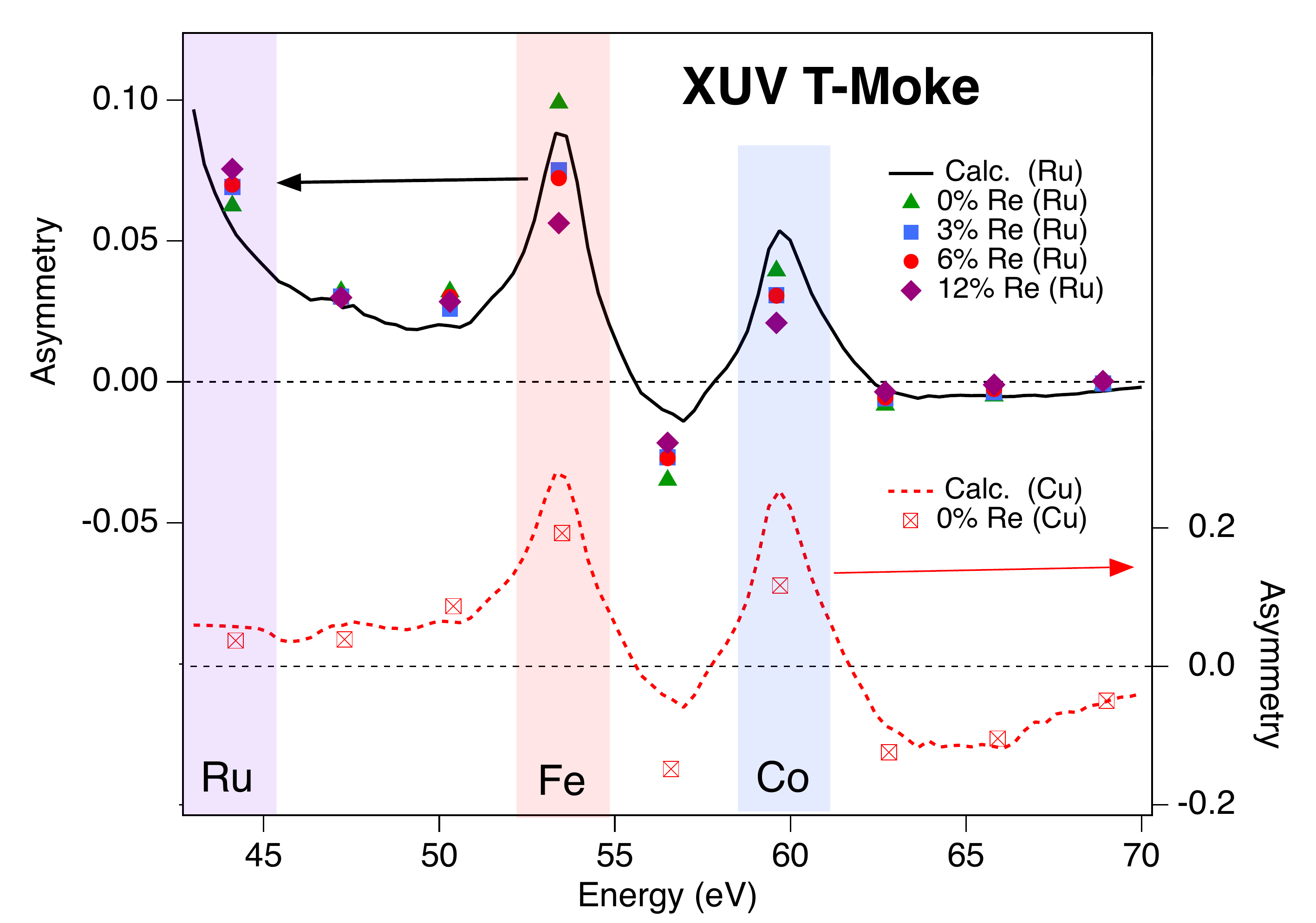}
    \caption{ Static T-MOKE asymmetries of  Ru/Fe$_{65}$Co$_{35}$/Ru heterostructures with different Re doping (left axis). The different shaded regions represent the absorption edges of Ru (purple), Fe (red), and Co (blue). The black solid line corresponds to the calculated asymmetry for the undoped sample. The right axis shows the asymmetry of an Cu/Fe$_{65}$Co$_{35}$/Cu heterostructure, both experimentally (red squares) and calculated (red dashed line). }
    \vspace*{-0.5cm}
    \label{Fig1}
\end{figure}


Figure \ref{Fig1} shows the measured static Kerr asymmetry of Ru/Fe$_{65}$Co$_{35}$/Ru for different Re concentrations. The asymmetry is $\sim$10\% and $\sim$4\% at the M$_{2,3}$-absorption edge of Fe ($\sim$54 eV) and Co ($\sim$60 eV), respectively. The asymmetry is reduced with increasing Re doping following the same trend as the saturation magnetization, which has  already been reported \cite{serkanenhaced,rahulenginner}. An additional large asymmetry is observed around 44 eV (
purple region), which corresponds to the Ru N$_{2,3}$-absorption edge. The magnitude of the Ru asymmetry is comparable to the Fe asymmetry at 54 eV. To verify that the increase in asymmetry at 44 eV is related to Ru, a Cu/Fe$_{65}$Co$_{35}$/Cu heterostructure was studied for which no significant increase in magnetic asymmetry near to 44 eV was observed (cf.\ lower part of Fig.\ \ref{Fig1}). Using the formalism given by Zak \textit{et al.}~\cite{ZAK.43.6423}, we have calculated 
the magnetic asymmetry for the undoped heterostructures and present them as black solid and red dashed lines in Fig.\ \ref{Fig1}, for the Ru and Cu capped samples, respectively. The magneto-optical (MO) parameters for Fe and Co were obtained from Willems \textit{et al.}~\cite{Willems_2020} and Valencia \textit{et al.}~\cite{Valencia_2006}. MO parameters are only available above 45 eV and was hence interpolated down to 44 eV (cf.\ Fig. S2, section I in SM \cite{SM}). Both Ru and Cu are non-magnetic in these simulations. From these simulations it is clear that the weak MO component from Fe around 44 eV will result in a large asymmetry at the Ru N$_{2,3}$ absorption edge, while for the Cu capped sample the asymmetry at 44 eV remains low. Characterization of the Ru interface utilizing element-specific synchrotron based x-ray techniques and first-principles calculations can be found in sections II and III of the SM \cite{SM}.

\begin{figure}[t!]
    \includegraphics[width=1.05\columnwidth]{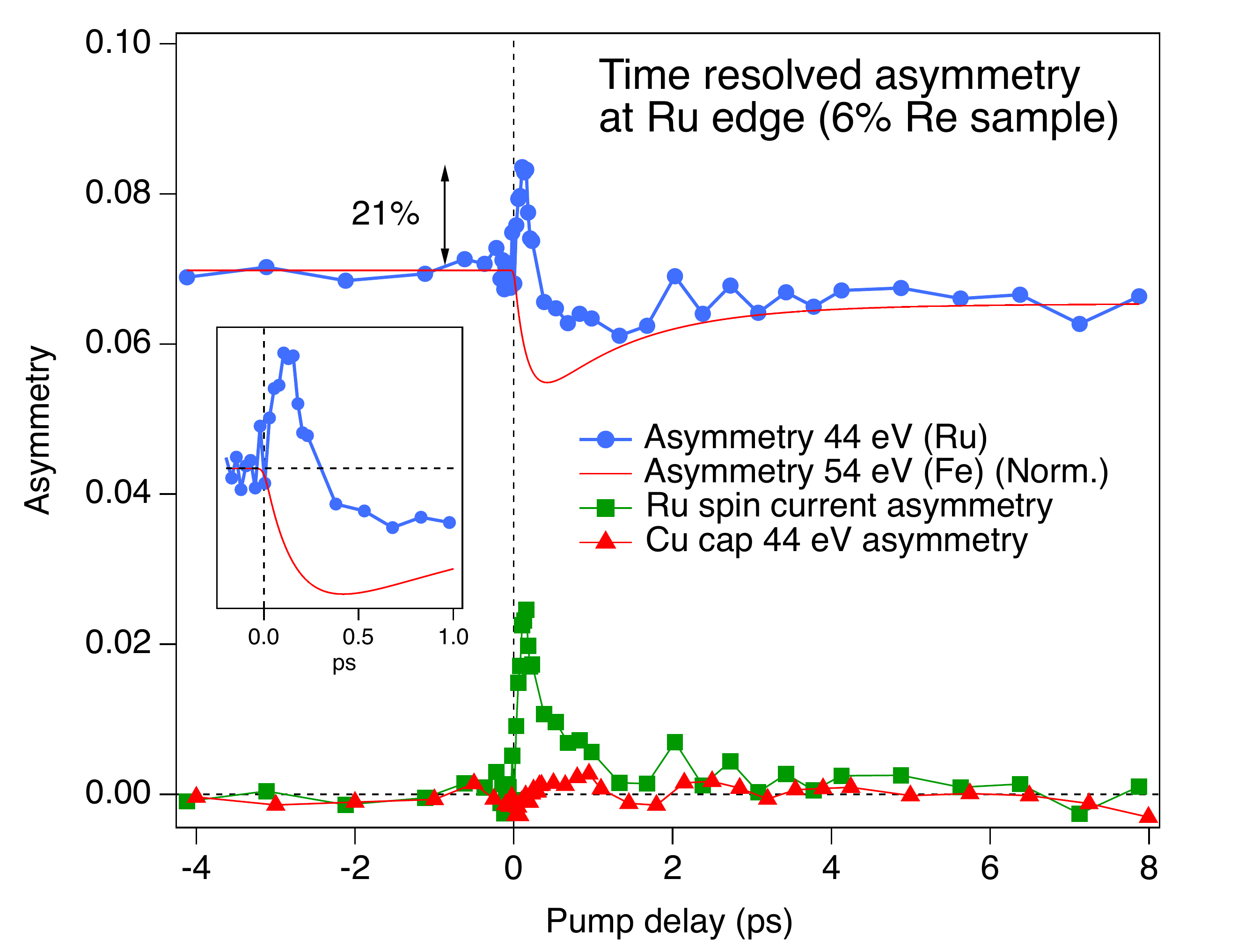}
	\caption{ The time-dependent magnetic asymmetry measured at the Ru edge shown with blue circles for the 6\% Re doped sample. The green squares correspond to the Ru asymmetry after subtracting the Fe contribution (red solid line).  
 The asymmetry at 44 eV for the Cu capped sample, after subtracting the Fe contribution is shown with red triangles. Inset shows the zoom view of Ru asymmetry.}
	 \vspace*{-0.4cm}
	\label{fig:Fig2_Ru_6pTrAsym}
\end{figure}

Figure\ \ref{fig:Fig2_Ru_6pTrAsym} shows time- and element-resolved measurements for the Ru capped 6\% Re doped sample. 
The blue circles shows the Ru asymmetry during the demagnetization. A large increase of the asymmetry ($\sim$ 21\%, as indicated in Figure\ \ref{fig:Fig2_Ru_6pTrAsym})  is observed after 100 fs, which has decreased to below its initial value after 500 fs. The initial rise of the asymmetry is attributed to spin currents that spin polarize Ru and the subsequent decreased asymmetry is related to the decreased asymmetry of Fe, which has a notable contribution at the Ru edge. The Fe contribution, represented by the fitted red solid line, can easily be subtracted from the Ru asymmetry at 44 eV. Note that the Fe asymmetry, measured at 54 eV, has been normalized to the Ru value before time zero. The remaining asymmetry then corresponds to the pure spin polarization of Ru, shown as green squares. It is clear that the peak of the Ru spin polarization occurs before the maximum demagnetization of Fe, and that this spin polarization is short lived ($\lesssim$\,500) fs. For comparison, we also show the asymmetry for the Cu capped sample at 44 eV after subtracting the Fe contribution (red triangles). It is clear that in this case there is no sign of any spin accumulation and that the asymmetry follows the Fe asymmetry (cf.\ Fig. S5, section I in SM \cite{SM}).

\begin{figure}
	\includegraphics[width=1\columnwidth]{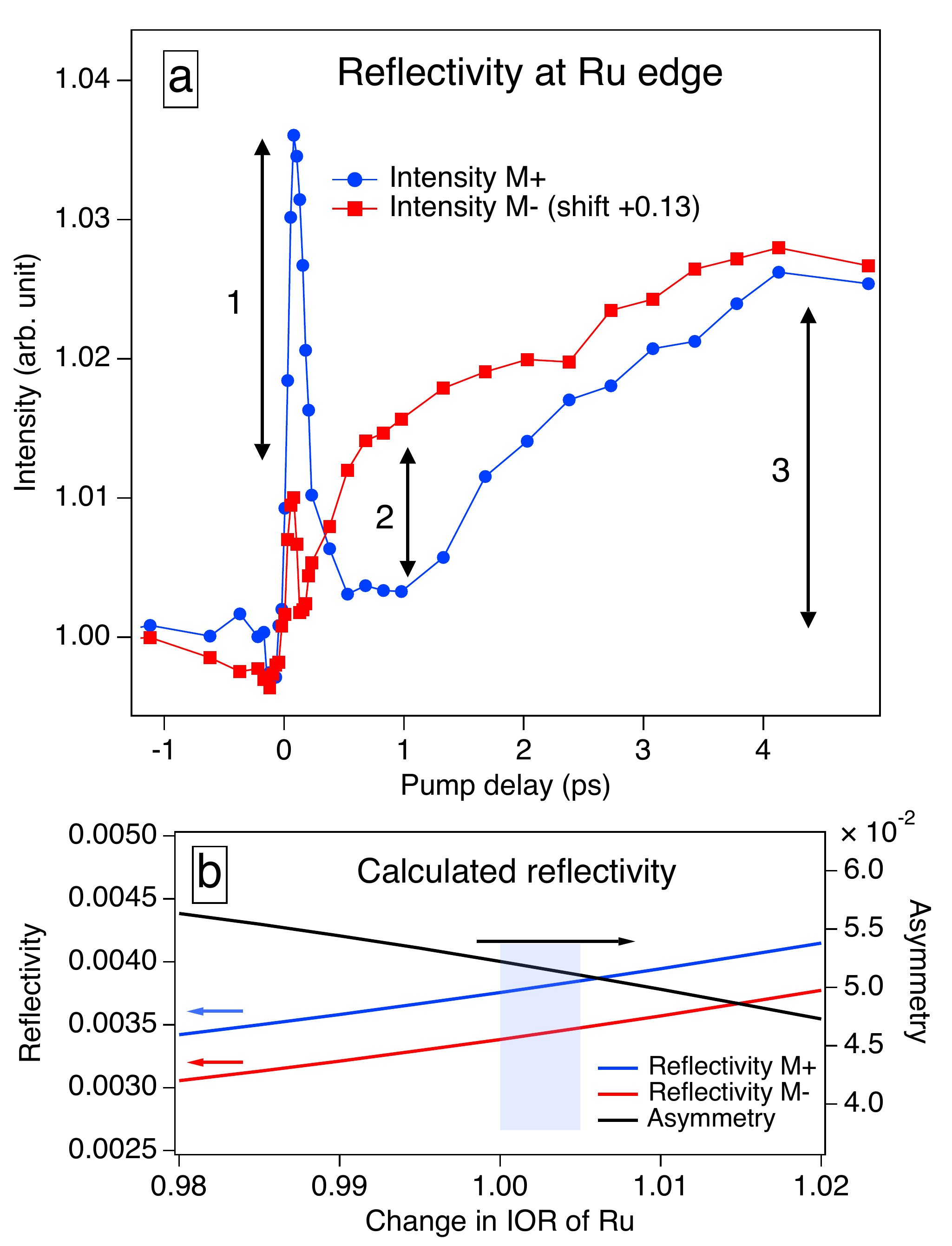}
	\caption{(a)  The time resolved reflectivity at the Ru edge (44 eV) for both magnetization directions ($M+$ and $M-$). 
At time zero there is a sharp increase in the reflectivity that is strongly asymmetric between $M+$ and $M-$ (cf. arrow 1).  After about 1 ps there is still a significant difference in reflectivity due to the demagnetization of the FeCo layer (cf. arrow 2). After about 4 ps there has been a slow increase in the reflectivity for both $M+$ and $M-$ (cf. arrow 3), suggesting that the index of refraction has changed. (b) Calculated dependence of the reflectivity (left axis, blue and red lines) and asymmetry (right axis, black line) on the index of refraction (IOR) of Ru. The transparent blue box indicates the experimentally found range of change in the reflectivity obtained from (a). 
}
	 \vspace*{-0.4cm}
	\label{fig:Ru_dynamics_Refl}
\end{figure}

It is possible that index of refraction of Ru at the absorption edge will be sensitive to absorption of the pump light and potentially create artefacts in the asymmetry signal \cite{Koopmans2000,Carva2009}.
To investigate this, we  show the reflectivity at the Ru edge (44 eV) for the two different magnetization directions in Fig.\ \ref{fig:Ru_dynamics_Refl}(a).
The data has been normalized to the value of $M+$ (blue circles) before time zero, after which the $M-$ data (red squares) has been shifted up by $0.13$ to facilitate the comparison between $M-$ and $M+$ data.
In Fig.\ \ref{fig:Ru_dynamics_Refl}(b) we show the calculated reflectivity for the two magnetization directions (blue and red lines, left scale) and the corresponding asymmetry (black line, right scale) as a function of the index of refraction (IOR) of Ru. The horizontal scale corresponds to a multiplicative factor in the IOR and $1.00$ hence corresponds to the unpumped case. The blue shaded box indicates the range that results in the same reflectivity changes as observed experimentally in  Fig.\ \ref{fig:Ru_dynamics_Refl}(a). Since changing the IOR has a similar effect on both $M+$ and $M-$, we find that the change in the observed asymmetry due changes in the IOR of Ru is fairly small. The asymmetry goes from 5.22\% to 5.11\% as one moves from 1.00 to 1.005 on the multiplication factor of IOR, i.e. left and right edges of the blue shaded box, and it should be noted that the asymmetry decreases rather than increases when the reflectivity increases. This would suggest that the spin current asymmetry in Fig.\ \ref{fig:Fig2_Ru_6pTrAsym} (green squares) is slightly underestimated, instead of peaking at about 2.5\% it should peak at 2.6\% if compensated for changes in the IOR of Ru. In Fig.\ \ref{fig:Ru_dynamics_Refl}(a) we observe an increase of both $M+$ and $M-$ intensities at time zero, suggesting a change in the IOR, however, there is a large difference in the amplitude of this increase (indicated by the vertical arrow denoted 1) which highlights the magnetic origin of the asymmetry change. The difference between $M+$ and $M-$ after 1 ps (vertical arrow denoted 2), is mainly due to the demagnetization of the FeCo layer. There is a slow increase in the reflectivity which saturates after 4 ps (vertical arrow denoted 3). This would suggest that the initial increase in reflectivity, at time zero, is due to a non-thermalized electron distribution. As the electron distribution thermalizes the reflectivity decreases, and finally the heat is transfered from the electron system to the phonon system, again resulting in an increased reflectivity.
\begin{figure}
	\includegraphics[width=1\columnwidth]{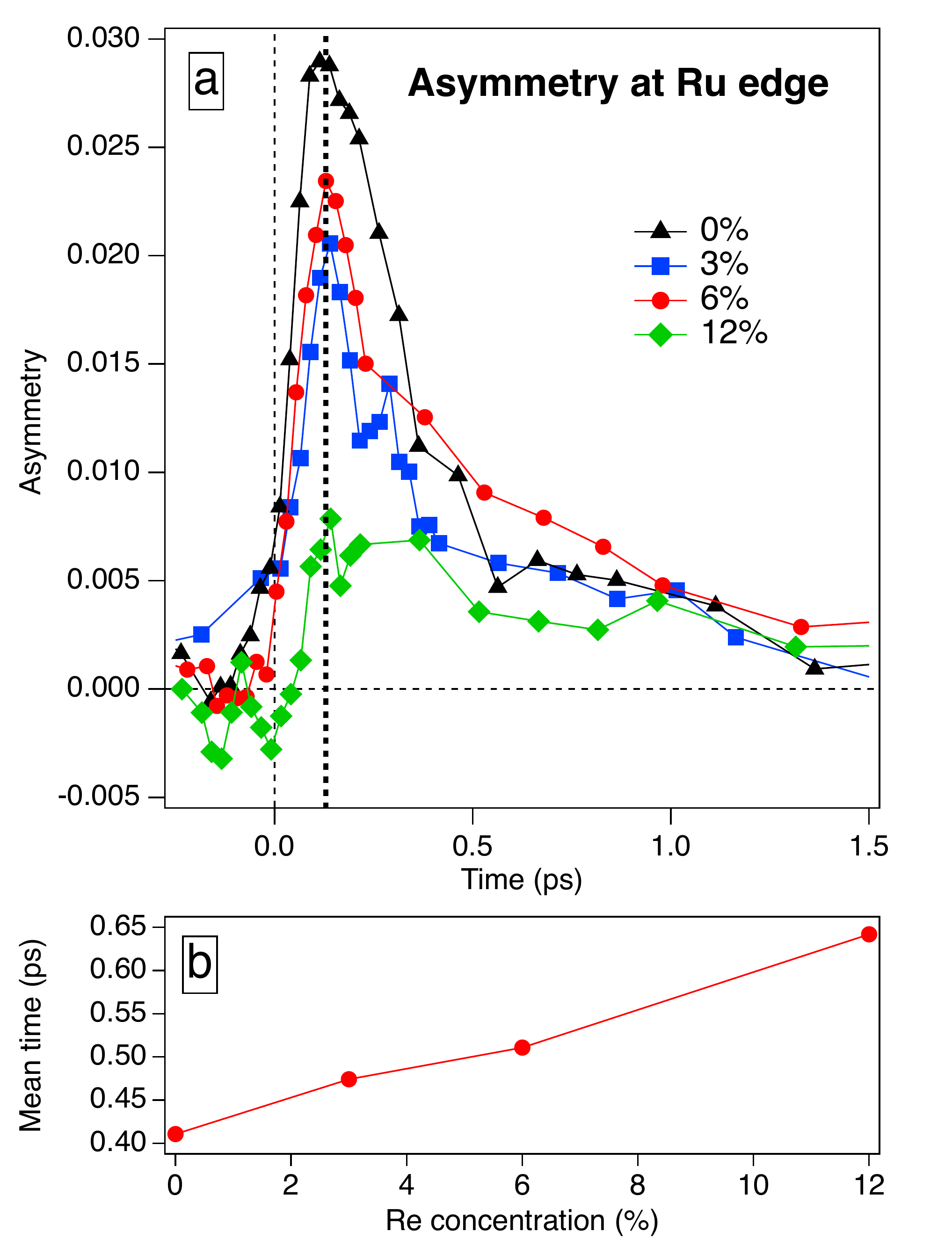}
	\caption{(a) Ultrafast magnetization dynamics of Ru (after subtracting the Fe contribution) in Fe$_{65}$Co$_{35}$ films with different Re doping. An increase in the asymmetry signal is observed   after time zero. The time-dependent change in asymmetry is strongest for the undoped sample and decreases with increasing Re doping. The vertical black dashed line provides a guide for the position of the Re 6\% sample peak position at 130 fs. (b) The extracted mean time of the experimentally observed asymmetry versus Re doping. The mean time of the asymmetry provides information about temporal shifts in the amplitude.  
	}
	 \vspace*{-0.4cm}
	\label{fig:Ru_dynamics_ReDoping}
\end{figure}


The ultrafast dynamics at the Ru absorption edge for different Re concentrations are shown in Fig.\ \ref{fig:Ru_dynamics_ReDoping}(a). In all measurements, the pump fluence was adjusted so that all samples showed a similar amount of demagnetization ($\sim$20\%) at the Fe edge. All curves have been smoothed by taking an average of each point by its nearest neighbouring points, cf. Fig.\ S6 in SM for non-smoothed data \cite{SM}. For the undoped sample, the peak asymmetry at the Ru edge is about { $\sim$3\%} (black curve). For the samples with 3 at\% and 6 at\% Re doping, the peak asymmetry signal decreases to $\sim$2\%.  Notably, Re doping attenuates the time dependent Ru asymmetry significantly, and for the 12.6 at.\% Re  doped sample the asymmetry is close to the detection limit. A reasonable explanation for this effect is that Re doping results in enhanced spin scattering, which effectively blocks the spin current from reaching the Ru layer as illustrated in Fig.\ S1(c) of the SM \cite{SM}. Also, the saturation magnetization decreases by 25\% from undoped to 12.6 \% Re doped samples \cite{rahulenginner}, which certainly should have a detrimental effect on the spin current. In Fig.\ \ref{fig:Ru_dynamics_ReDoping}(b) we show the mean time of the asymmetry amplitude (i.e. mean time $=\sum_{t=0}^{t=1.5 ps} A(t)\cdot t /\sum_{t=0}^{t=1.5 ps} A(t) $, where $A(t)$ is the time dependent asymmetry). Since the asymmetry at shorter time scales $\sim$ 130 fs appears to be affected more by the Re doping than the asymmetry after 0.5 ps, the mean time shifts to higher values with increasing Re doping. 

\begin{figure}
	\includegraphics[width=1\columnwidth]{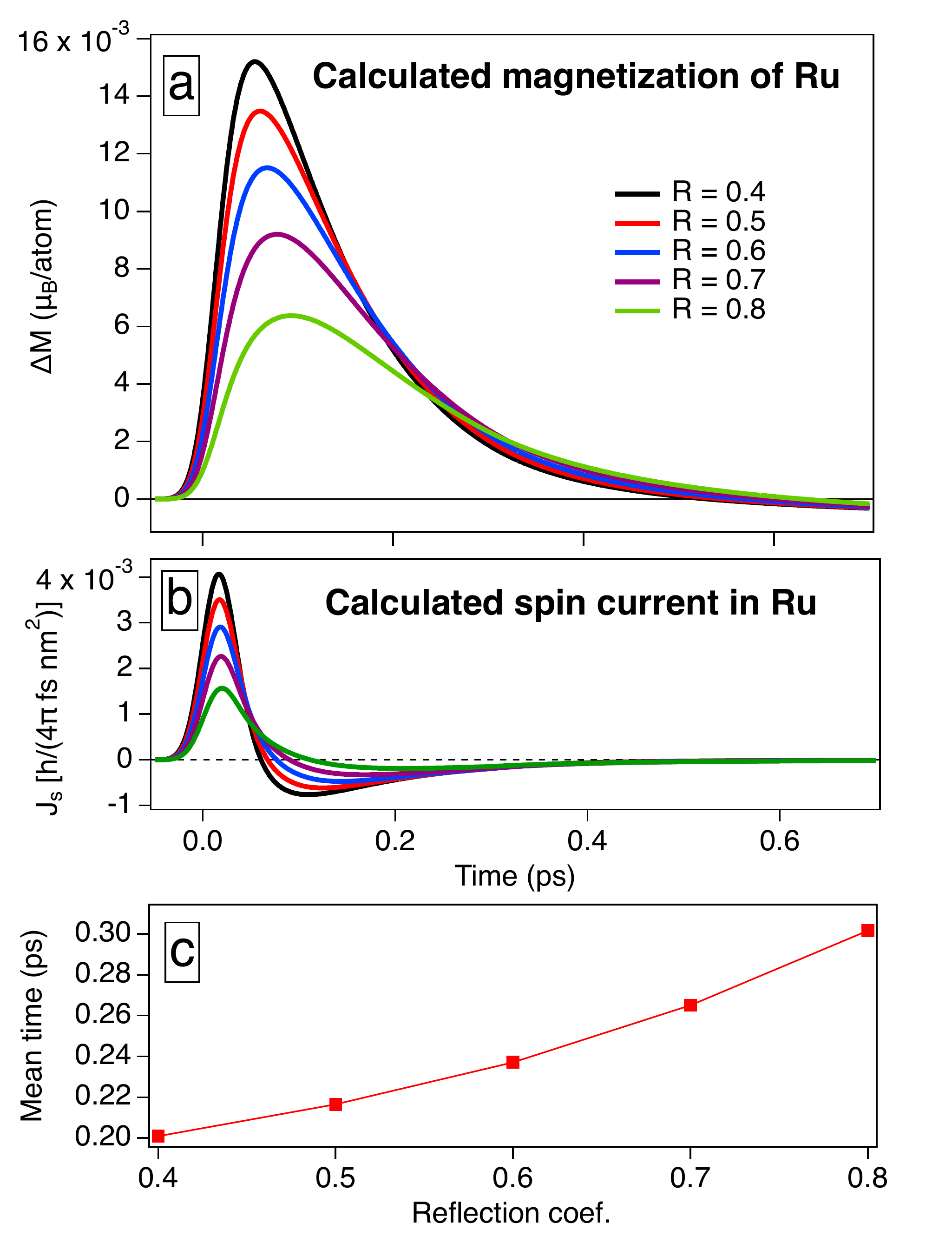}
	\caption{ (a) Computed time-dependent magnetization in Ru for diﬀerent reﬂection coeﬃcients at the Ru/FeCo interface. (b) Calculated spin current in Ru for different reﬂection coeﬃcients at the Ru/FeCo interface. (c) The extracted mean time of the simulated asymmetry versus reflection coefficient. The mean time of the asymmetry amplitude moves to higher values with an increasing reflection coefficient. } 
	 \vspace*{-0.4cm}
	\label{fig:Ru_dynamics_theory}
\end{figure}

In Figure\ \ref{fig:Ru_dynamics_theory} we show the calculated superdiffusive spin current in the Ru layer using the superdiffusive spin transport model. 
 In this model the laser-induced demagnetization of FeCo is the source of the spin current \cite{Battiato_2010}.
The details of the model can be found in section IV, SM \cite{SM}.
Figure\ \ref{fig:Ru_dynamics_theory}(a) displays the computed laser-induced transient magnetization of the Ru layer as a function of the spin current reflectivity at the FeCo/Ru interface, assuming an Fe$_{0.5}$Co$_{0.5}$ composition and using \textit{ab initio} computed hot electron velocities and spin lifetimes from Ref.\ \cite{Nechaev2009}. The  magnetization of the  Ru layer consists of two contributions to properly treat the interface,  $\Delta M_{\mathrm {Ru}}(3  \, \mathrm {nm})$ and a small fraction of the magnetization of the  ferromagnetic layer, $\Delta M_{\mathrm {FeCo}}(0.2 \,  \mathrm {nm})$. This is done because in the superdiffusive model  material properties change abruptly from one layer to the other without taking into account hybridization and other interface effects. We notice that the Ru layer reaches its maximum magnetization at $t  \sim  60 $ fs, where $t = 0$ is taken at the peak of the gaussian laser  pulse. After this fast magnetization, the layer is demagnetized within 0.6 ps. Note that the magnetization becomes slightly negative due to the demagnetization of the $\Delta M_{\mathrm {FeCo}}(0.2 \,  \mathrm {nm})$ layer that was included at the interface. In Figure\ \ref{fig:Ru_dynamics_theory}(b) we show the calculated spin current in Ru. Around the peak Ru magnetization the spin current changes its direction and flows back into the FeCo layer. Except for the trivial reduction in spin current amplitude when increasing the reflection coefficient, since less spins will transmit through the interface, we also see a significant reduction in the spin current that flows back into the FeCo layer. Effectively the increased reflection coefficient traps the spins more effectively within the Ru layer, and the reduction in Ru magnetization will hence be increasingly more driven by spin-flip processes rather than by the back flow. In Figure\ \ref{fig:Ru_dynamics_theory}(c) we have calculated the mean time of the magnetization in the range between 0 and 0.6 ps, similarly as the mean time was calculated for the asymmetry in Fig.\ \ref{fig:Ru_dynamics_ReDoping}(b). We find that an increased reflectivity at the interface results in a similar behaviour as we find experimentally when increasing the Re doping.

The present measurements are consistent with superdiffusive spin-current injection into Ru, but other models based on spin injection through magnon spin pumping have also been proposed \cite{Choi2015,Beens2020,Lichtenberg2022}. In the former, the spin current is generated through the difference in transport efficiency between majority and minority spins. In the latter, mainly majority electrons are generated during demagnetization via electron-magnon interactions, which creates a spin imbalance and a resulting spin-current. Interestingly, the spin-mixing conductance \cite{PhysRevLett.88.117601} of the FeCo/Ru interface has been found to increase with Re doping \cite{rahulenginner}, suggesting that the strong electron energy dependence of the reflectivity \cite{LuZB+2020} can result in an enhanced spin-mixing conductance (low energy electrons) at the same time as the
superdiffusive spin current (high energy electrons) decreases due to higher reflectivity. The Gilbert damping parameter ($\alpha_G$) is known to increase when transition-metal films are doped with 4d, 5d or rare-earth elements \cite{Rantschler_2007,  Woltersdorf_2009, Rebei_2006, akansel2018effect}.
For our samples, $\alpha_G$ increases more than four-fold for 12.6\% Re doping in FeCo \cite{rahulenginner}. A magnon dominated demagnetization process would likely be sensitive to the Gilbert damping \cite{Koopmans_2005}, however, we find no change of the demagnetization time for the different samples (cf. Fig.\ S4(a), SM \cite{SM}). Furthermore, in models based on spin pumping through magnon damping \cite{Choi2015,Beens2020,Lichtenberg2022} more Gilbert damping would imply an increased loss of spin angular momentum and a spin voltage imbalance building up at the FeCo/Ru interface, leading to a larger spin current pumped into Ru. However, this behavior is not seen in our measurements.   

We expect that other non-magnetic (NM) transition metals should exhibit a similar contrast at their respective M or L-edges, if they are used as capping layers. For buried NM layers it is unlikely that the signal can be separated from the much stronger signal resulting from the ferromagnetic layer. 
Previously, observation of laser-excited spin currents were made with visible MOKE \cite{Hofherr2017,Alekhin2019} and magnetic second-harmonic generation \cite{Melnikov2011}. Recently, injection of nonequilibrium electrons from Fe into Au were measured with time-resolved photoemission \cite{PhysRevResearch.4.033239}. However, direct element-selective measurements, as presented here, of spin injection in NM metals due to superdiffusive
spin currents have not been reported before.
These results are of particular use for studying devices that emit THz radiation generated from spin currents \cite{kampfrath2013terahertz}, a phenomenon that has received attention recently \cite{Nenno2019,Dang2020,gupta2021co2feal,Papaioannou2021,Bull2021,gupta2021strain,mottamchetty2023direct}. Figure S10 of the SM depicts the Fourier transform of the superdiffussive THz spin current in the Ru layer; the bandwidth is found to be $ \sim 18$ THz for the undoped sample, which decreases with increasing Re doping. 

This work is supported by the Swedish Research Council (VR, contracts 2017-03799, 2021-04658 and 2021-5395), the Olle Engkvists Stiftelse (Grant No.\ 182-0365), Carl Tryggers Stiftelse (Grant No.\ 17:241), and by the Swedish National Infrastructure for Computing (SNIC), through grant agreement No.\ 2018-05973. This work is furthermore funded by the European Union's Horizon2020 Research and Innovation Programme under FET-OPEN Grant agreement No.\ 863155 (s-Nebula). P.M.O.\ acknowledges support by the Deutsche Forschungsgemeinschaft through CRC/TRR 227 ``Ultrafast Spin Dynamics", project MF. R.G.\ acknowledges DESY (Hamburg, Germany) for the provision of experimental facilities. Parts of this research were carried out at beamline P09 at PETRA III under proposal I-20200894 EC. We are grateful to M.Sc.\ Dominik Graulich and Dr.\ Timo Kuschel (Bielefeld University) for developing within the long-term project LTP I-20170016 the vacuum chamber for the vector electromagnet used in this work and Prof.\ G{\"u}nter Reiss (Bielefeld University) for contributing within LTP II-20190009 to the in-vacuum silicon drift detector used in this work.


\bibliographystyle{apsrev4-1}
\bibliography{apssamp}

\end{document}